\documentclass[%
 reprint,
superscriptaddress,
 amsmath,amssymb,
 aps,
]{revtex4-2}

\usepackage{graphicx}
\usepackage{dcolumn}
\usepackage{upgreek}
\usepackage{bm}

\usepackage[showframe],

\begin{document}

\title{Programmable on-chip synthesis and reconstruction of partially coherent two-mode optical fields}

\author{Amin Hashemi}
\affiliation{CREOL, The College of Optics \& Photonics, University of Central Florida, Orlando, FL 32816, USA}

\author{Abbas Shiri}
\affiliation{CREOL, The College of Optics \& Photonics, University of Central Florida, Orlando, FL 32816, USA}

\author{Bahaa E. A. Saleh}
\affiliation{CREOL, The College of Optics \& Photonics, University of Central Florida, Orlando, FL 32816, USA}

\author{Andrea Blanco-Redondo}
\affiliation{CREOL, The College of Optics \& Photonics, University of Central Florida, Orlando, FL 32816, USA}

\author{Ayman F. Abouraddy}
\affiliation{CREOL, The College of Optics \& Photonics, University of Central Florida, Orlando, FL 32816, USA}
\affiliation{raddy@creol.ucf.edu}




\begin{abstract}
Partially coherent light is typically studied in the context of freely propagating continuous fields. Recent developments have indicated the existence of a `coherence advantage' in multimode optical communications, where partially coherent light outperforms coherent light. However, exploiting partial coherence in such applications requires manipulating multimode field coherence in programmable on-chip platforms. We present here the first example of on-chip synthesis and characterization of two-mode optical fields in an integrated on-chip hexagonal mesh of Mach-Zehnder interferometers. Starting with incoherent two-mode light, we adjust the degree of coherence on the chip with non-unitary transformations, construct $2\times2$ unitary transformations to synthesize prescribed coherence matrices, and reconstruct the coherence matrices via measurements of the spatial Stokes parameters. These results indicate the possibility of deploying programmable photonics for producing large-dimensional structured partially coherent light for applications in communications, cryptography, sensing, and spectroscopy. 
\end{abstract}

\maketitle
\section{Introduction}

All natural sources of light are partially coherent, and the study of coherence has been a mainstay of optical physics since Michelson's experiments in astronomy \cite{Michelson1891PM1} and Zernike's quantification of coherence via the visibility of interference fringes \cite{Zernicke}. Emil Wolf and others established since the 1950's a rigorous foundation for optical coherence based on continuous correlation functions in space and time \cite{Wolf54NCC,Wolf55PRSL,Wolf59INC,Karczewski63INC,Mandel65RMP,Perina72Book,Born99Book,Wolf07Book,Goodman15Book,Agarwal20PO}. To date, the province of investigating the dynamics of partial coherence remains primarily that of freely propagating fields \cite{Gbur10PO,Korotkova20PO} (with few exceptions in optical fibers \cite{Saleh81AO} and surface plasmon polaritons \cite{Norman15OE,Norrman16EL,Chen19PRA,Chen20PO}), with applications in light propagation in turbulent atmospheres \cite{Gbur02JOSAA,Ponomarenko02OL,Dogariu03OL,Shirai03JOSAA}, solar energy \cite{Divitt15Optica,Ricketti22SR}, near-field thermal radiation \cite{Yu23PRL}, among a host of other possibilities \cite{Korotkova20PO}.

Two recent developments motivate a departure from this well-trodden path. First, what can be called the `coherence advantage' has been recently identified in several applications involving optical information processing \cite{Nardi22OL,Dong24Nature,Harling25APLP}. By `coherence advantage' we refer to scenarios in which partially coherent light outperforms coherent light in an application previously restricted to coherent light. A well-known example is speckle reduction in imaging when utilizing partially coherent light rather than coherent light \cite{Fujii75NRO,Goodman07Book,Peng21SA,Evered25OLT}. More recently, the coherence advantage has emerged in communications applications making use of structured fields comprising a finite number of modes \cite{Nardi22OL,Harling25APLP}. For example, it has been suggested theoretically that partially coherent light over a communications link supporting $N$ modes (e.g., a multimode fiber) provides $\mathcal{O}(N^{2})$ independent channels encoded in the correlations between the modes (corresponding to the off-diagonal elements of the $N\times N$ coherence matrix), rather that the conventional $\mathcal{O}(N)$ channels available with coherent light (the diagonal elements of the coherence matrix) \cite{Nardi22OL}. More recently, it has been demonstrated experimentally that the coherence rank \cite{Harling24PRA2} (the number of non-zero eigenvalues of the coherence matrix) associated with partially coherent light can be exploited as a scattering-immune information carrier even across a strongly scattering channel \cite{Harling25APLP}. Such experiments have been carried out to date in free space with bulk optics at slow speeds. Bringing these capabilities closer to real applications requires transitioning to integrated on-chip platforms.

Second, the general trend in both classical \cite{Bogaerts20Nature,Zhang23APL,Shekhar24NC} and quantum optics \cite{Politi09IEEEJSTQE,Wang20NP,Elshaari20NP,Labonte24PRX} over the past decade has been to exploit high-speed, compact-footprint photonic integrated circuits in lieu of bulk optics. Indeed, integrated-photonics platforms alone provide the requisite stability in settings that comprise a large number of cascaded interferometers \cite{Miller17OE,Pai19PRA,Bell21APLP}. However, only coherent light so far has been exploited in programmable photonics, so that partially coherent light has not yet taken advantage of these novel opportunities. Rather than remaining the purview of freely propagating fields, the study of partial coherence can benefit from the transition to programmable on-chip platforms in novel fundamental studies of entropy swapping and concentration \cite{Okoro17Optica,Harling22OE,Harling23JO}, locked entropy \cite{Harling24PRA}, and behaviors that depend on the coherence rank \cite{Harling24PRA2} (e.g., interference visibility in presence of multiple physical degrees of freedom \cite{Abouraddy17OE} and optical cross-purity \cite{Mandel61JOSA}), in addition to potential applications of partially coherent fields in optical information processing. 

Here we present the first experimental demonstration of on-chip manipulation of partially coherent optical fields. We make use of fields comprising two spatial modes (corresponding to the fields in a pair of single-mode waveguides), and exploit an on-chip platform comprising a large hexagonal mesh of programmable Mach-Zehnder interferometers (MZIs) to achieve the following tasks: (1) to construct $2\times2$ \textit{non-unitary} transformations that tune the degree of coherence starting with generic incoherent light; (2) to construct $2\times2$ \textit{unitary} transformations that mold the partially coherent field in a diagonal representation to yield a prescribed coherence matrix; and (3) sequential implementation of software-controlled $2\times2$ unitary transformations for the reconstruction of the spatial coherence matrix via measurements of the spatial Stokes parameters utilizing on-chip detectors. We thus establish experimentally for the first time on-chip synthesis and characterization of two-mode partially coherent light. Because the unitary transformations reported here are the fundamental building blocks of their larger-dimensional counterparts \cite{Reck94PRL,Zukowski97PRA,Bogaerts20Nature}, these results indicate conclusively the potential for on-chip programmable manipulation of `structured coherence' for applications in optical communications \cite{Harling25APLP}, cryptography \cite{Peng21P,Liu25LPR}, computation \cite{Dong24Nature}, and spectroscopy \cite{Miller25Optica}.

\section{Partially coherent two-mode fields}

\subsection{The coherence matrix for a two-mode field}

\begin{figure}[t!]
\centering
\includegraphics[width=8.6cm]{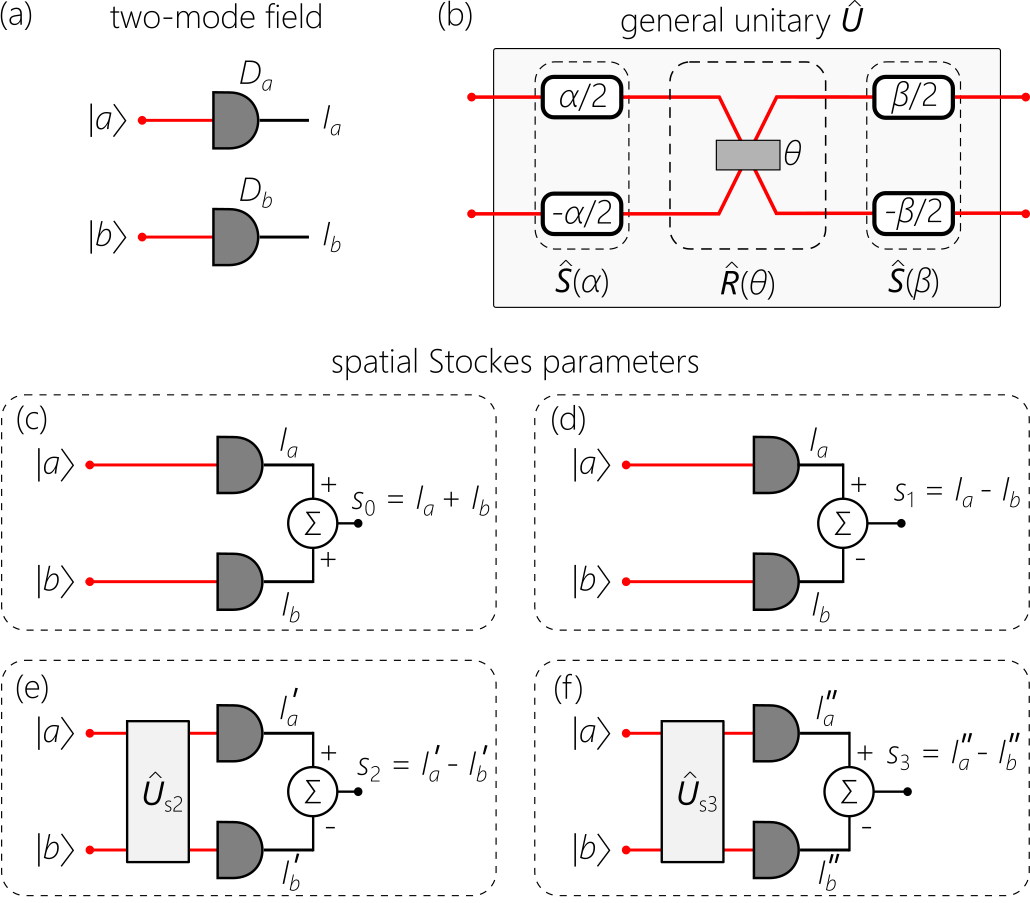} 
\caption{(a) Two-mode optical field comprising the modes $|a\rangle$ and $|b\rangle$ in single-mode waveguides. Two detectors obtain the modal weights: $I_{a}=G_{aa}$ and $I_{b}=G_{bb}$. (b) An arbitrary unitary $\hat{U}$ (Eq.~\ref{eq:GeneralUnitary}) is constructed from a phase operator $\hat{S}(\alpha)$, followed by a variable coupler $\hat{R}(\theta)$, and then a second phase operator $\hat{S}(\beta)$. (c-f) Configurations to measure the spatial Stokes parameters and thus reconstruct the coherence matrix: (c) $s_{0}=I_{a}+I_{b}$; (d) $s_{1}=I_{a}-I_{b}$; (e) $s_{2}=I_{a}'-I_{b}'$ after the unitary $\hat{U}_{s2}$; and (f) $s_{3}=I_{a}''-I_{b}''$ after the unitary $\hat{U}_{s3}$.}
\label{fig:TheoryUnitary}
\end{figure}

We consider a scalar optical field comprising two spatial modes denoted $|a\rangle$ and $|b\rangle$ in the Dirac notation, which are here the fields in a pair of single-mode optical waveguides [Fig.~\ref{fig:TheoryUnitary}(a)]. When the field is coherent, it can be described by a field vector,
\begin{equation}
|E\rangle=\left(\begin{array}{c}E_{a}\\E_{b}\end{array}\right)=E_{a}\left(\begin{array}{c}1\\0\end{array}\right)+E_{b}\left(\begin{array}{c}0\\1\end{array}\right)=E_{a}|a\rangle+E_{b}|b\rangle,    
\end{equation}
where we use matrix representations for $|a\rangle$ and $|b\rangle$, $E_{a}$ and $E_{b}$ are the complex modal amplitudes, and we normalize the field vector to $\langle E|E\rangle=1$, so that $|E_{a}|^{2}+|E_{b}|^{2}=1$ \cite{Gamo64PO}.

A partially coherent field is represented by a $2\times2$ Hermitian, unity-trace, positive semi-definite spatial coherence matrix \cite{Gamo64PO}:
\begin{equation}\label{eq:CoherenceMatrix}
\mathbf{G}=\left(\begin{array}{cc}G_{aa}&G_{ab}\\G_{ba}&G_{bb}\end{array}\right),
\end{equation}
where $G_{jk}=\langle E_{j}E_{k}^{*}\rangle$, $\langle\cdot\rangle$ is an ensemble average, $G_{ba}=G_{ab}^{*}$, and the trace is $\mathrm{Tr}\{\mathbf{G}\}=G_{aa}+G_{bb}=1$. We define the degree of spatial coherence as:
\begin{equation}
D_{\mathrm{s}}=\lambda_{a}-\lambda_{b}=\sqrt{1-4\mathrm{det}\{\mathbf{G}\}},
\end{equation}
where $1\geq\lambda_{a}\geq\lambda_{b}\geq0$ are the two eigenvalues of $\mathbf{G}$, $\lambda_{a}+\lambda_{b}=1$, and $\mathrm{det}\{\mathbf{G}\}=G_{aa}G_{bb}-|G_{ab}|^{2}$ is the determinant of $\mathbf{G}$. This is the usual definition of the degree of polarization used in optics \cite{Brosseau06PO} applied here to the spatial degree-of-freedom \cite{Kagalwala13NP,Abouraddy17OE,Eberly17Optica,Abouraddy19Optica,Halder21OL}, which and corresponds to the maximum achievable visibility of spatial interference \cite{Zernicke}. The treatment of partially coherent fields in a discrete modal basis has a long history \cite{Gamo64PO} , but has only been utilized for the discretization of freely propagating fields.

The diagonal elements $G_{aa}=\langle E_{a}E_{a}^{*}\rangle$ and $G_{bb}=\langle E_{b}E_{b}^{*}\rangle$ of $\mathbf{G}$ represent the fractions of the total power associated with modes $|a\rangle$ and $|b\rangle$, respectively, and can thus be estimated by simply measuring the modal weights or the optical power in each waveguide [Fig.~\ref{fig:TheoryUnitary}(a)]. The off-diagonal element $G_{ab}=\langle E_{a}E_{b}^{*}\rangle$ cannot be measured directly in this way. When the two modes are superposed, the visibility of the interference fringes is $V=2|G_{ab}|\leq D_{\mathrm{s}}$ and the phase of $G_{ab}$ causes a shift in the fringes, with $V=D_{\mathrm{s}}$ reached only when $G_{aa}=G_{bb}=\tfrac{1}{2}$ \cite{Zernicke,Abouraddy19Optica}.

\subsection{Two-mode unitary transformations}

Applications in information processing require the implementation of general unitary transformations (`unitaries' for brevity), represented here by $2\times2$ unitary matrices $\hat{U}$ on the space of $2\times2$ spatial coherence matrices; here $\hat{U}\hat{U}^{\dagger}=\hat{\mathbb{I}}_{2}$ and $\mathrm{det}\{\hat{U}\}=1$, where $\hat{\mathbb{I}}_{2}$ is the $2\times2$ identity matrix. The most general $2\times2$ unitary $\hat{U}$ with $\mathrm{det}\{\hat{U}\}=1$ takes the form:
\begin{equation}\label{eq:GeneralUnitary}
\hat{U}=\left(\begin{array}{cc}
e^{i\varphi_{1}}\cos\tfrac{\theta}{2}&
-e^{-i\varphi_{2}}\sin\tfrac{\theta}{2}\\
e^{i\varphi_{2}}\sin\tfrac{\theta}{2}&
e^{-i\varphi_{1}}\cos\tfrac{\theta}{2}
\end{array}\right),    
\end{equation}
which can be decomposed into a sequence of three simpler unitaries [Fig.~\ref{fig:TheoryUnitary}(b)]: a variable rotator $\hat{R}(\theta)$  sandwiched between two phase operators $\hat{S}(\alpha)$ and $\hat{S}(\beta)$ that introduce phase shifts $\alpha$ and $\beta$ between the two modes, respectively, where:
\begin{equation}\label{eq:RotatorPhase}
\hat{R}(\theta)=\left(\begin{array}{cc}\cos\tfrac{\theta}{2}&-\sin\tfrac{\theta}{2}\\\sin\tfrac{\theta}{2}&\cos\tfrac{\theta}{2}\end{array}\right),\;\hat{S}(\alpha)=\left(\begin{array}{cc}e^{i\alpha/2}&0\\0&e^{-i\alpha/2}\end{array}\right).
\end{equation}
The general unitary in Eq.~\ref{eq:GeneralUnitary} is then given by $\hat{U}=\hat{S}(\beta)\;\hat{R}(\theta)\;\hat{S}(\alpha)$. after setting $\alpha=\varphi_{1}+\varphi_{2}$ and $\beta=\varphi_{1}-\varphi_{2}$.

\subsection{Reconstructing two-mode coherence matrices}

The spatial coherence matrix in Eq.~\ref{eq:CoherenceMatrix} can be expanded in terms of the Pauli matrices $\{\hat{\sigma}_{j}\}_{j=0}^{3}$:
\begin{equation}\label{eq:GinTermsOfStokes}
\mathbf{G}=\tfrac{1}{2}\sum_{j=0}^{3}s_{j}\hat{\sigma}_{j}=\tfrac{1}{2}\left(\begin{array}{cc}s_{0}+s_{1}&s_{2}-is_{3}\\s_{2}+is_{3}&s_{0}-s_{1}\end{array}\right),
\end{equation}
where the expansion coefficients $s_{j}=\mathrm{Tr}\{\hat{\sigma}_{j}\mathbf{G}\}$ correspond to the Stokes parameters applied here to the spatial degree-of-freedom of the field rather than to polarization as is more familiar, $\hat{\sigma}_{0}=\hat{\mathbb{I}}_{2}$, and:
\begin{equation}
\hat{\sigma}_{1}=\left(\begin{array}{cc}1&0\\0&-1\end{array}\right),\;
\hat{\sigma}_{2}=\left(\begin{array}{cc}0&1\\1&0\end{array}\right),\;
\hat{\sigma}_{3}=\left(\begin{array}{cc}0&-i\\i&0\end{array}\right).
\end{equation}
The spatial Stokes parameters are real because $\mathbf{G}$ is Hermitian, $s_{0}=1$ because $\mathbf{G}$ is unity-trace, $|s_{j}|\leq1$, and $D_{\mathrm{s}}=\sqrt{s_{1}^{2}+s_{2}^{2}+s_{3}^{2}}$.

The spatial Stokes parameters can be measured using the four configurations depicted in Fig.~\ref{fig:TheoryUnitary}(c-f): $s_{0}=G_{aa}+G_{bb}=I_{a}+I_{b}$ obtained by adding the modal weights [Fig.~\ref{fig:TheoryUnitary}(c)]; $s_{1}=G_{aa}-G_{bb}=I_{a}-I_{b}$ by subtracting the modal weights [Fig.~\ref{fig:TheoryUnitary}(d)]; $s_{2}=2\mathrm{Re}\{G_{ab}\}=I_{a}'-I_{b}'$ is the difference between the modal weights after traversing the unitary $\hat{U}_{\mathrm{s}2}=\tfrac{1}{\sqrt{2}}\left(\begin{array}{cc}1&-1\\1&1\end{array}\right)$ [Fig.~\ref{fig:TheoryUnitary}(e)]; and $s_{3}=-2\mathrm{Im}\{G_{ab}\}=I_{a}''-I_{b}''$ is the difference between the modal weights after traversing the unitary $\hat{U}_{\mathrm{s}3}=\tfrac{1}{\sqrt{2}}\left(\begin{array}{cc}1&i\\i&1\end{array}\right)$ [Fig.~\ref{fig:TheoryUnitary}(f)]. Substituting the measured Stokes parameters in Eq.~\ref{eq:GinTermsOfStokes} reconstructs the coherence matrix \cite{Abouraddy14OL,Kagalwala15SR}.

\section{On-chip implementation of unitaries}

\subsection{Two-mode on-chip unitaries}

\begin{figure}[t!]
\centering
\includegraphics[width=8.7cm]{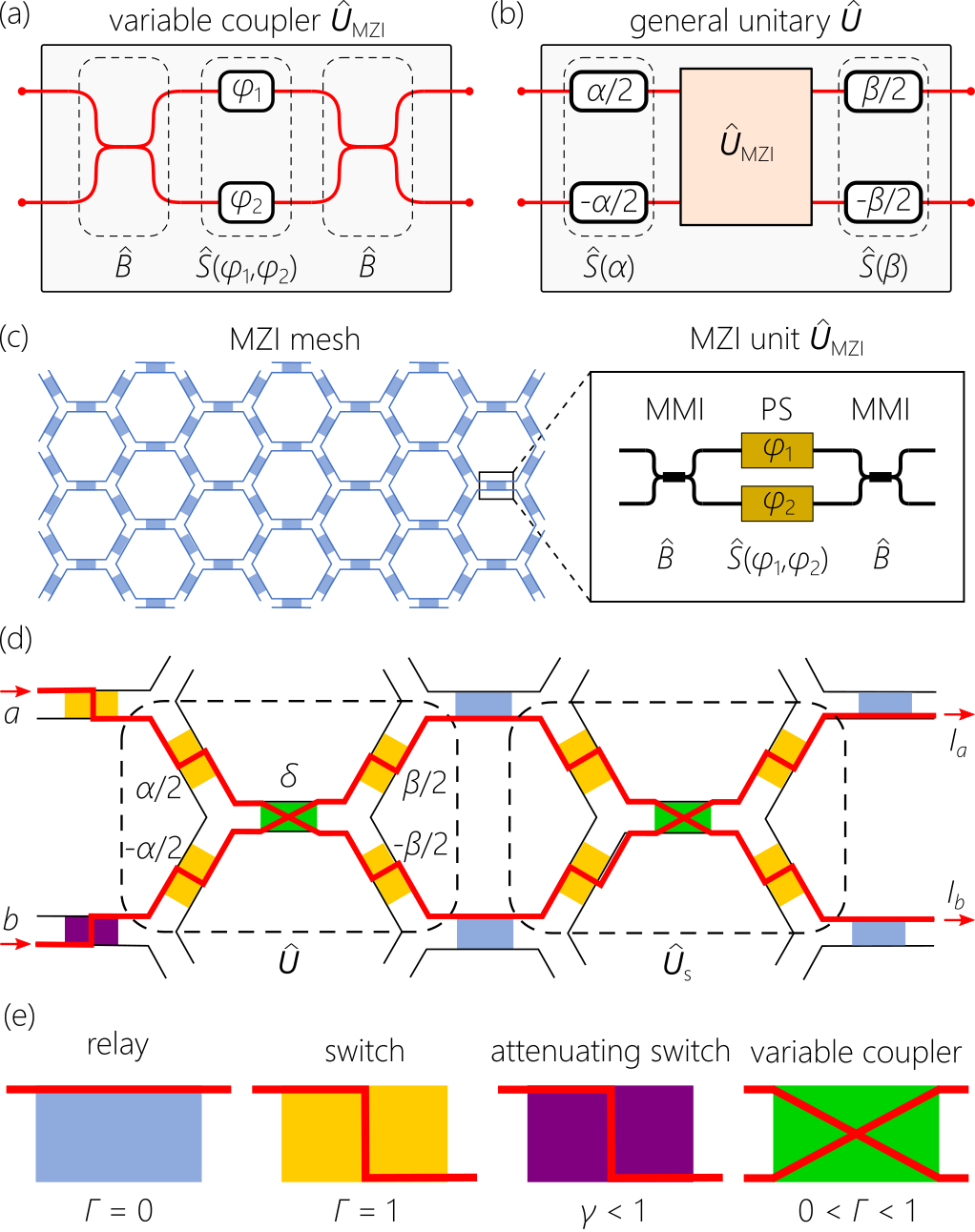} 
\caption{(a) A variable coupler $\hat{U}_{\mathrm{MZI}}$ (Eq.~\ref{eq:MZI}) is constructed by sandwiching a phase operator $\hat{S}(\varphi_{1},\varphi_{2})$ between two symmetric couplers represented by the operator $\hat{B}$. (b) A general on-chip unitary (Eq.~\ref{eq:GeneralOnChip}) is constructed by sandwiching a variable coupler $\hat{U}_{\mathrm{MZI}}$ between two phase operators $\hat{S}(\alpha)$ and $\hat{S}(\beta)$. (c) Schematic of the on-chip hexagonal mesh utilized in our experiments. Each of the~72 colored rectangles in the mesh is an MZI, and the lines are single-mode waveguides. The inset highlights the structure of an MZI, which corresponds to the MZI in (a). (d) The layout of the MZIs used to construct a general unitary $\hat{U}$, followed by the Stokes unitary $\hat{U}_{\mathrm{s}}$. At the entrance to $\hat{U}$ is a \textit{non}-unitary transformation that adjusts the degree of spatial coherence of the diagonal coherence matrix delivered to the unitary. The outputs are then routed to detectors at the edge of the chip to record the modal weights $I_{a}$ and $I_{b}$. (e) A key to the MZIs of different coupling parameter $\Gamma$ depicted in different colors in (d). From left to right: a relay with $\Gamma=0$ corresponds to an MZI that routes $|a\rangle\rightarrow|a\rangle$ and $|b\rangle\rightarrow|b\rangle$; a switch with $\Gamma=1$ that routes $|a\rangle\rightarrow|b\rangle$ and $|b\rangle\rightarrow|a\rangle$; an attenuating switch that reduces the amplitude of the field $|a\rangle\rightarrow\gamma|b\rangle$, with $\gamma<1$, and there is no input field at $|b\rangle$; and, finally, a variable coupler given by Eq.~\ref{eq:MZI}.}
\label{fig:Chip}
\end{figure}

On-chip implementations place restrictions on the realizable elements. Specifically, it is challenging to directly implement a variable coupler $\hat{R}(\theta)$ between two waveguides as a single on-chip element. However, a readily available phase operator $\hat{S}(\varphi_{1},\varphi_{2})$ that introduces phases $\varphi_{1}$ and $\varphi_{2}$ to the modes $|a\rangle$ and $|b\rangle$, respectively, sandwiched between two fixed symmetric couplers $\hat{B}=\tfrac{1}{\sqrt{2}}\left(\begin{array}{cc}1&i\\i&1\end{array}\right)$ yields:
\begin{equation}\label{eq:MZI}
\hat{U}_{\mathrm{MZI}}(\delta,\varphi)=\hat{B}\;\hat{S}(\varphi_{1},\varphi_{2})\;\hat{B}=ie^{i\varphi}\left(\begin{array}{cc}\sin\frac{\delta}{2}&\cos\frac{\delta}{2}\\\cos\frac{\delta}{2}&-\sin\frac{\delta}{2}\end{array}\right),
\end{equation}
thus corresponding to a variable coupler [Fig.~\ref{fig:Chip}(a)]; here $\delta=\varphi_{1}-\varphi_{2}$ determines the coupling ratio between the two modes and $\varphi=\tfrac{1}{2}(\varphi_{1}+\varphi_{2})$ is an overall phase. By adding phase operators $\hat{S}(\alpha)$ and $\hat{S}(\beta)$ (Eq.~\ref{eq:RotatorPhase}) before and after the MZI, we obtain the general unitary:
\begin{equation}\label{eq:GeneralOnChip}
\hat{U}=ie^{i\varphi}\left(\begin{array}{cc}
e^{i\xi_{1}}\sin\frac{\delta}{2}&e^{-i\xi_{2}}\cos\frac{\delta}{2}\\e^{i\xi_{2}}\cos\frac{\delta}{2}&-e^{-i\xi_{1}}\sin\frac{\delta}{2}\end{array}\right),
\end{equation}
where $\xi_{1}=\tfrac{1}{2}(\alpha+\beta)$ and $\xi_{2}=\tfrac{1}{2}(\alpha-\beta)$ [Fig.~\ref{fig:Chip}(b)].

\subsection{Programmable on-chip platform}

The programmable silicon-photonic chip used in our experiments (iPronics Smartlight Processor) consists of the hexagonal mesh of 72~MZIs depicted in Fig.~\ref{fig:Chip}(c), with each MZI formed of two phase shifters sandwiched between a pair of $2\times2$ multimode interferometers [Fig.~\ref{fig:Chip}(c), inset]. The physical length of each MZI is $\approx811$~$\mu$m, with an average insertion loss of $\approx0.5$~dB (which does not affect the implementation of the unitaries). The chip provides 28~accessible optical channels at its perimeter, which serve as configurable input and output ports for injecting or extracting optical signals. The average fiber-to-chip coupling loss is 3.34~dB, in addition to on-chip routing losses of approximately $6-8$~dB to and from the optical input/output ports. Optical power is measured using integrated photodetectors (PDs) with a sensitivity of -70~dBm. All input/output ports, MZIs, and PDs are fully configurable and are controlled through a Python-based software
development kit.

The construction of a general unitary makes use of a section of the hexagonal MZI mesh as depicted in Fig.~\ref{fig:Chip}(d). Two input ports are selected on the left to correspond to modes $|a\rangle$ and $|b\rangle$. Each mode can be routed from the input port to the location of the unitary. It is useful to consider first some on-chip building blocks used throughout this work. Starting with the unitary representing the MZI (Eq.~\ref{eq:MZI}), we define a coupling coefficient $\Gamma=\cos^{2}\tfrac{\delta}{2}$, which corresponds to the strength of coupling between modes $|a\rangle$ and $|b\rangle$, $\hat{U}_{\mathrm{MZI}}=ie^{i\varphi}\left(\begin{array}{cc}\sqrt{1-\Gamma}&\sqrt{\Gamma}\\\sqrt{\Gamma}&-\sqrt{1-\Gamma}\end{array}\right)$. Several special cases are identified in Fig.~\ref{fig:Chip}(e). When $\Gamma=0$, we have $\hat{U}_{\mathrm{MZI}}=ie^{i\varphi}\left(\begin{array}{cc}1&0\\0&-1\end{array}\right)$, and there is no exchange between the fields, which we refer to as a `relay'. When $\Gamma=1$, we have $\hat{U}_{\mathrm{MZI}}=ie^{i\varphi}\left(\begin{array}{cc}0&1\\1&0\end{array}\right)$, so that the fields are fully exchanged between $|a\rangle$ and $|b\rangle$, which we refer to as a `switch'. We also make use of an `attenuating switch' comprising an MZI with no field directed to one of the input ports. The input field at the other port is directed to the output with a power attenuation factor $\gamma=\Gamma<1$. The remaining input power that emerges from the other output port is discarded. Otherwise, the MZI represents a `variable' coupler with coupling strength $0<\Gamma<1$.  

\begin{figure}[t!]
\centering
\includegraphics[width=8.7cm]{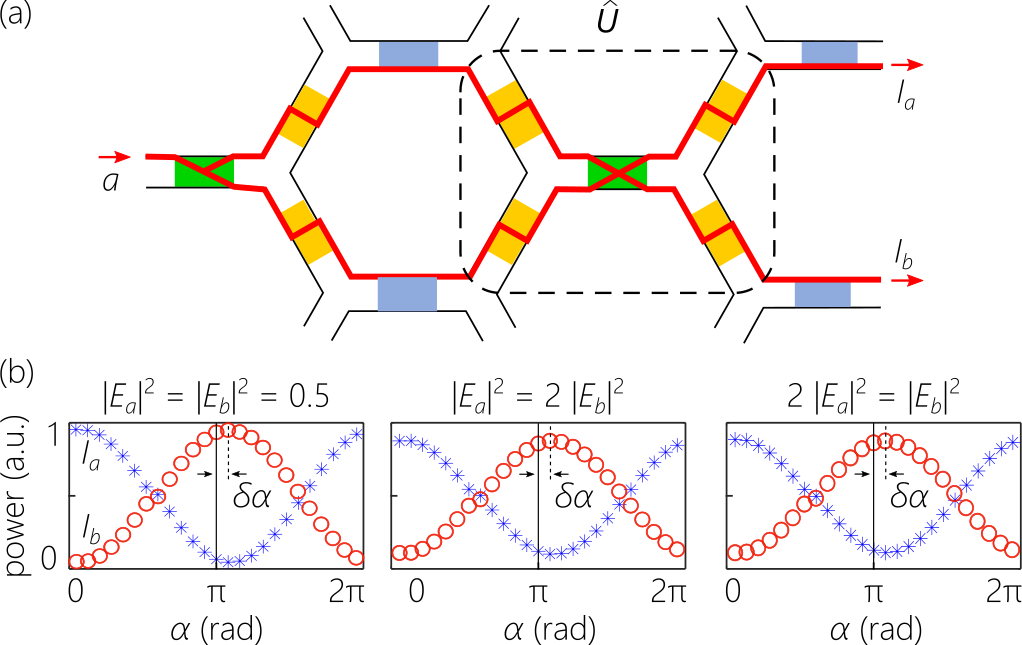} 
\caption{(a) The on-chip configuration to calibrate any relative on-chip phases between $|a\rangle$ and $|b\rangle$ en route to $\hat{U}$. Preceding $\hat{U}$ is a variable coupler that splits a coherent input to two paths with different relative amplitudes, so that $|E\rangle=E_{a}|a\rangle+E_{b}|b\rangle$. (b) Measurement results for $I_{a}$ and $I_{b}$ to calibrate the phase offset between $|a\rangle$ and $|b\rangle$ from the input ports to $\hat{U}$. The phase $\alpha$ in $\hat{U}$ is tuned for three different settings of the modal weights.}
\label{fig:Calibration}
\end{figure}

In our experiments we start with generic incoherent light $\mathbf{G}=\tfrac{1}{2}\hat{\mathbb{I}}_{2}$ ($D_{\mathrm{s}}=0$) coupled to the chip. We route the modes $|a\rangle$ and $|b\rangle$ from the chip entrance to the section depicted in Fig.~\ref{fig:Chip}(d). First, a \textit{non}-unitary transformation is implemented that is represented by the operator $\hat{\Lambda}=\left(\begin{array}{cc}1&0\\0&\sqrt{\gamma}\end{array}\right)$ to tune the degree of spatial coherence by transforming the coherence matrix from $\mathbf{G}=\tfrac{1}{2}\hat{\mathbb{I}}_{2}$ ($D_{\mathrm{s}}=0$) to $\mathbf{G}=\tfrac{1}{1+\gamma}\left(\begin{array}{cc}1&0\\0&\gamma\end{array}\right)=\left(\begin{array}{cc}\lambda_{a}&0\\0&\lambda_{b}\end{array}\right)$ with $D_{\mathrm{s}}=\tfrac{1-\gamma}{1+\gamma}$. Once the target degree of spatial coherence is realized, the fields traverse the unitary $\hat{U}$ to synthesize the prescribed coherence matrix.

The unitary $\hat{U}$ is followed by a second unitary $\hat{U}_{\mathrm{s}}$ (referred to as the Stokes unitary) to reconstruct the coherence matrix synthesized by $\hat{U}$. The modes $|a\rangle$ and $|b\rangle$ are then routed to detectors at the edge of the chip to measure the modal weights $I_{a}$ and $I_{b}$. 

However, there may exist an unspecified relative phase encountered by the signals along the waveguides before reaching the unitary. Because the on-chip operations that we implement are phase-sensitive, it is important that we calibrate for such phases and compensate for them. We perform this calibration by using an on-chip laser diode (wavelength $\approx1.55$~$\mu$m) coupled internally to port $|a\rangle$ at the input, while no input is provided to port $|b\rangle$ [Fig.~\ref{fig:Calibration}(a)], corresponding to a coherent vector field $|E\rangle=\left(\begin{array}{c}1\\0\end{array}\right)$. We direct $|E\rangle$ to a variable coupler that prepares the field vector $|E\rangle=\left(\begin{array}{c}|E_{a}|\\|E_{b}|e^{i\delta\alpha}\end{array}\right)$, where $\delta\alpha$ is the unknown phase when arriving at $\hat{U}$. The output from $\hat{U}$ is directed to the on-chip detectors to record $I_{a}$ and $I_{b}$. We tune the phase $\alpha$ in $\hat{U}$ (setting $\beta=0$ and $\delta=\tfrac{\pi}{2}$) for three cases of the input field amplitudes: $|E_{a}|^{2}=|E_{b}|^{2}=\tfrac{1}{2}$, $|E_{a}|^{2}=2|E_{b}|^{2}$, and $2|E_{a}|^{2}=|E_{b}|^{2}$. We plot in Fig.~\ref{fig:Calibration}(b) the measurements of $I_{a}$ and $I_{b}$ in these cases, from which we estimate an offset phase difference of $\delta\alpha\approx0.13\pi$, which we compensate for hereon. 

\section{Experiments}

\begin{figure}[t!]
\centering
\includegraphics[width=8.7cm]{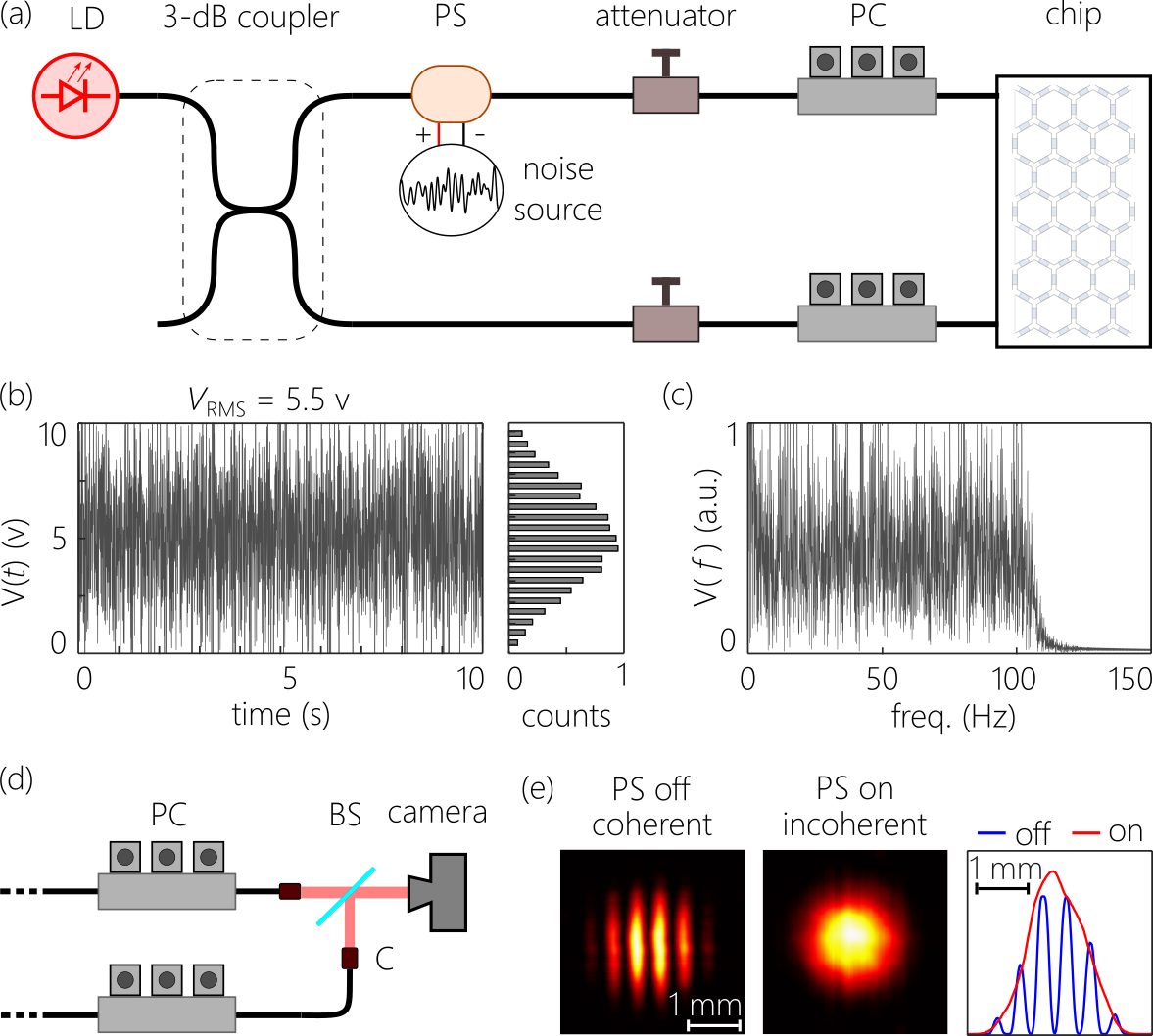} 
\caption{(a) Schematic of the source of incoherent light used in our experiments. LD: Laser diode; PS: phase shifter; PC: polarization controller. Black lines are all single-mode fibers. (b) Measured temporal noise signal with amplitudes from~0 to~10~v. In the right panel we plot a histogram of the noise signal showing a root-mean-square width $\approx5.5$~v. (d) Spectrum of the noise signal, showing a cutoff at $\approx100$~Hz. (d) Setup for observing the interference fringes produced by overlapping the two fields in free space prior to coupling to the chip. BS: Symmetric beam splitter; C: fiber collimator. (e) Images acquired by the camera of the intensity resulting from superposing the two fields with a slight angular tilt. When the phase shifter is turned off, high-visibility fringes are observed ($V\approx1$; left panel). When the phase shifter is fed with the noise signal from (b), the fringes are eliminated, signifying loss of coherence ($V\approx0$, middle panel). In the right panel we overlap sections through the two intensity distributions.}
\label{fig:Source}
\end{figure}

\subsection{Optical source}

The setup to prepare incoherent light is shown in Fig.~\ref{fig:Source}(a). It is challenging to efficiently couple two mutually incoherent fields into two single-mode fibers (SMFs), which will then couple to the chip. We make use of a laser diode (CoBrite DX1, IDPHOTONICS) at a wavelength $\approx1.55$~$\mu$m, linewidth $<25$~kHz, and 10-mW of power. The laser is split into two SMFs via a 3-dB coupler, with one output directed to a phase shifter (FPS-002-L, General Photonics). The two SMFs are then connected to the chip after variable fiber attenuators to balance the powers in the two SMFs and polarization controllers to match the polarization in the SMF to the chip for efficient coupling [Fig.~\ref{fig:Source}(a)]. We use a function generator (SIGLENT, SDG 1062X) to provide a noise signal to the phase shifter [Fig.~\ref{fig:Source}(b)]. A phase shift of $0-2\pi$ corresponds to $0-5$~v. The Fourier transform of the noise signal confirms the target bandwidth of $\approx100$~Hz [Fig.~\ref{fig:Source}(c)].

We confirm the incoherence of this two-mode source by examining the visibility $V$ of the spatial interference fringes resulting from superposing the two modes in free space. Before coupling the SMFs to the chip, we out-couple the two fields into free space through collimators (F220APC-1550, Thorlabs). Each collimator yields a beam of width $\approx2$~mm, which are combined at a symmetric beam splitter (BS015, Thorlabs) and directed to a camera (Bobcat-320-GigE-13907, Xenics) placed $\approx30$~cm away from the fiber collimators (averaging time $\approx50$~ms); see Fig.~\ref{fig:Source}(d). The When the phase shifter is turned off, the two modes are mutually \textit{coherent}, and interference fringes with $V\approx1$ are observed [Fig.~\ref{fig:Source}(e)]. When the noise signal drives the phase shifter, the two modes are mutually \textit{incoherent}, and no fringes are observed [Fig.~\ref{fig:Source}(e)]. Because the amplitudes at $|a\rangle$ and $|b\rangle$ are equal, the coherence matrix is diagonal $\mathbf{G}=\tfrac{1}{2}\hat{\mathbb{I}}_{2}$ when the phase shifter is fed the noise signal, and otherwise is $\mathbf{G}=\tfrac{1}{2}\left(\begin{array}{cc}1&1\\1&1\end{array}\right)$, representing a coherent field.

\begin{figure*}[t!]
\centering
\includegraphics[width=17.5cm]{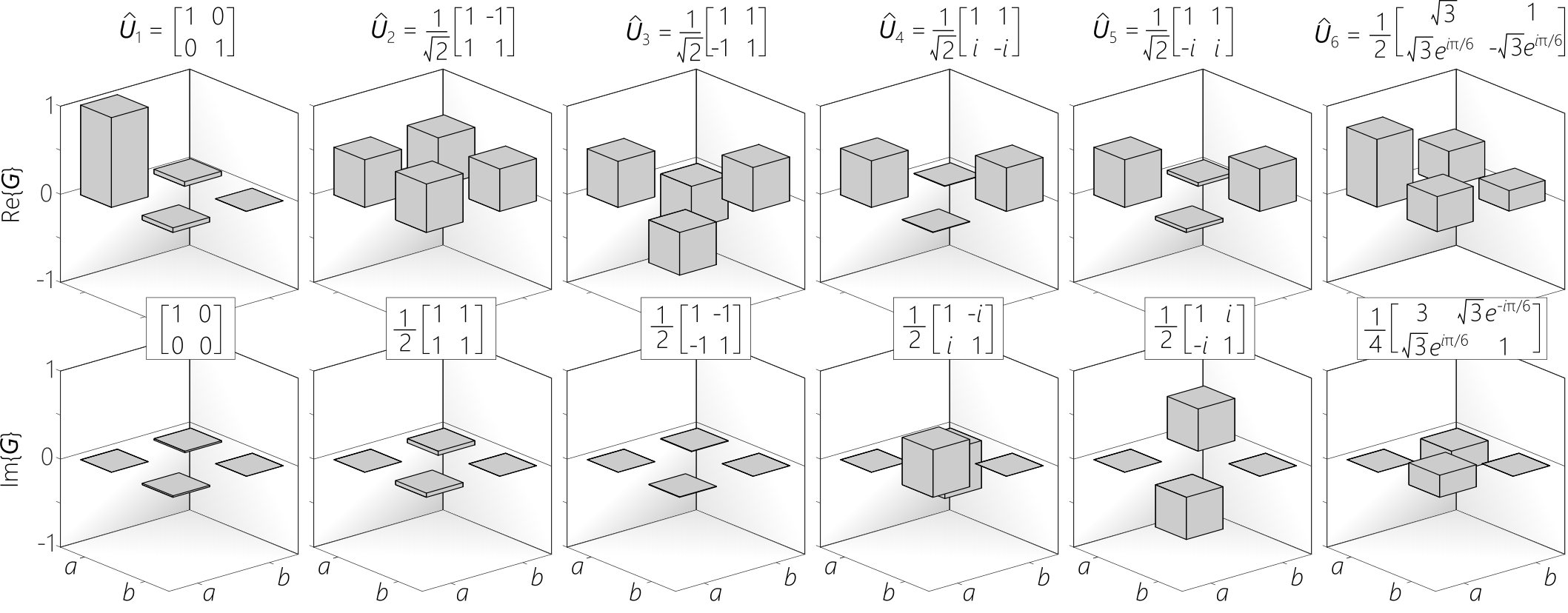} 
\caption{Reconstruction of the spatial coherence matrix $\mathbf{G}=\hat{U}\mathbf{G}^{\mathrm{D}}\hat{U}^{\dagger}$ for a coherent field, $\mathbf{G}^{\mathrm{D}}=\left(\begin{array}{cc}1&0\\0&0\end{array}\right)=|E\rangle\langle E|$, with $|E\rangle=\left(\begin{array}{c}1\\0\end{array}\right)$. The columns correspond to $\mathbf{G}$ after the unitaries provided at the top, which are those given in Eq.~\ref{eq:6Unitaries} after dropping the overall phases, and the expected $\mathbf{G}$ is provided in each column. We plot the real and imaginary parts of the coherence matrices, $\mathrm{Re}\{\mathbf{G}\}$ and $\mathrm{Im}\{\mathbf{G}\}$, respectively.}
\label{fig:CoherentData}
\end{figure*}

We make use of the source in our experiments with the noise signal fed to the phase shifter, so that the chip is provided with $\mathbf{G}=\tfrac{1}{2}\hat{\mathbb{I}}_{2}$. The on-chip non-unitary transformation tunes the degree of spatial coherence and produces the diagonal coherence matrix $\mathbf{G}^{\mathrm{D}}=\left(\begin{array}{cc}\lambda_{a}&0\\0&\lambda_{b}\end{array}\right)$, and the unitary $\hat{U}$ produces the new coherence matrix:
\begin{equation}\label{eq:GeneralHermitian}
\mathbf{G}=\hat{U}\mathbf{G}^{\mathrm{D}}\hat{U}^{\dagger}=\frac{1}{2}\left(\begin{array}{cc}
1-D_{\mathrm{s}}\cos\delta&D_{\mathrm{s}}e^{-i\alpha}\sin\delta\\
D_{\mathrm{s}}e^{-i\alpha}\sin\delta&1+D_{\mathrm{s}}\cos\delta
\end{array}\right).
\end{equation}
Note that setting $\delta=\tfrac{\pi}{2}$ guarantees that the diagonal elements are equalized and the magnitude of the off-diagonal element is $\tfrac{1}{2}D_{\mathrm{s}}$, which will be useful below.

\subsection{Reconstruction of the coherence matrix via the spatial Stokes parameters}

To reconstruct the coherence matrix, we make use of a second unitary that we refer to as the Stokes unitary $\hat{U}_{\mathrm{s}}$ [Fig.~\ref{fig:Chip}(d)] to implement the sequence of measurements depicted in Fig.~\ref{fig:TheoryUnitary}(c-f) and estimate the spatial Stokes parameters. We sequentially implement on the chip three settings to acquire the Stokes parameters. In each setting, a new coherence matrix is produced $\mathbf{G}'=\hat{U}_{\mathrm{s}}\mathbf{G}\hat{U}_{\mathrm{s}}^{\dagger}=\left(\begin{array}{cc}G_{aa}'&G_{ab}'\\G_{ba}'&G_{bb}\end{array}\right)$, and we recode the modal weights $I_{a}=G_{aa}'$ and $I_{b}=G_{bb}'$ via on-chip detectors. The elements of $\mathbf{G}'$ are related to the sought-after elements of $\mathbf{G}$ via $\hat{U}_{\mathrm{s}}$.

First, to acquire the Stokes parameters $s_{0}$ and $s_{1}$, we set $\delta=\pi$ in Eq.~\ref{eq:GeneralOnChip} to obtain $\hat{U}_{\mathrm{s}}=ie^{i\varphi}\left(\begin{array}{cc}1&0\\0&-1\end{array}\right)$, in which case $I_{a}=G_{aa}'=G_{aa}$ and $I_{b}=G_{bb}'=G_{bb}$, so that $s_{0}=I_{a}+I_{b}=G_{aa}+G_{bb}$ and $s_{1}=I_{a}-I_{b}=G_{aa}-G_{bb}$. Next, we set $\delta=\tfrac{\pi}{2}$ and $\alpha=0$ to implement $\hat{U}_{\mathrm{s}2}=ie^{i\varphi}\tfrac{1}{\sqrt{2}}\left(\begin{array}{cc}1&1\\1&-1\end{array}\right)$, in which case $s_{2}=I_{a}'-I_{b}'=2\mathrm{Re}\{G_{ab}\}$. Finally, we set $\delta=\tfrac{\pi}{2}$ and $\alpha=\tfrac{\pi}{2}$ to implement $\hat{U}_{\mathrm{s}3}=ie^{i\varphi}\tfrac{1}{\sqrt{2}}\left(\begin{array}{cc}1&-i\\i&-1\end{array}\right)$, in which case $s_{2}=I_{a}'-I_{b}'=-2\mathrm{Im}\{G_{ab}\}$. These settings enable the acquisition of the spatial Stokes parameters $\{s_{j}\}$, from which we reconstruct the coherence matrix $\mathbf{G}$ (Eq.~\ref{eq:GinTermsOfStokes}).

\subsection{Measurements for a coherent field}

\begin{figure*}[t!]
\centering
\includegraphics[width=15.85cm]{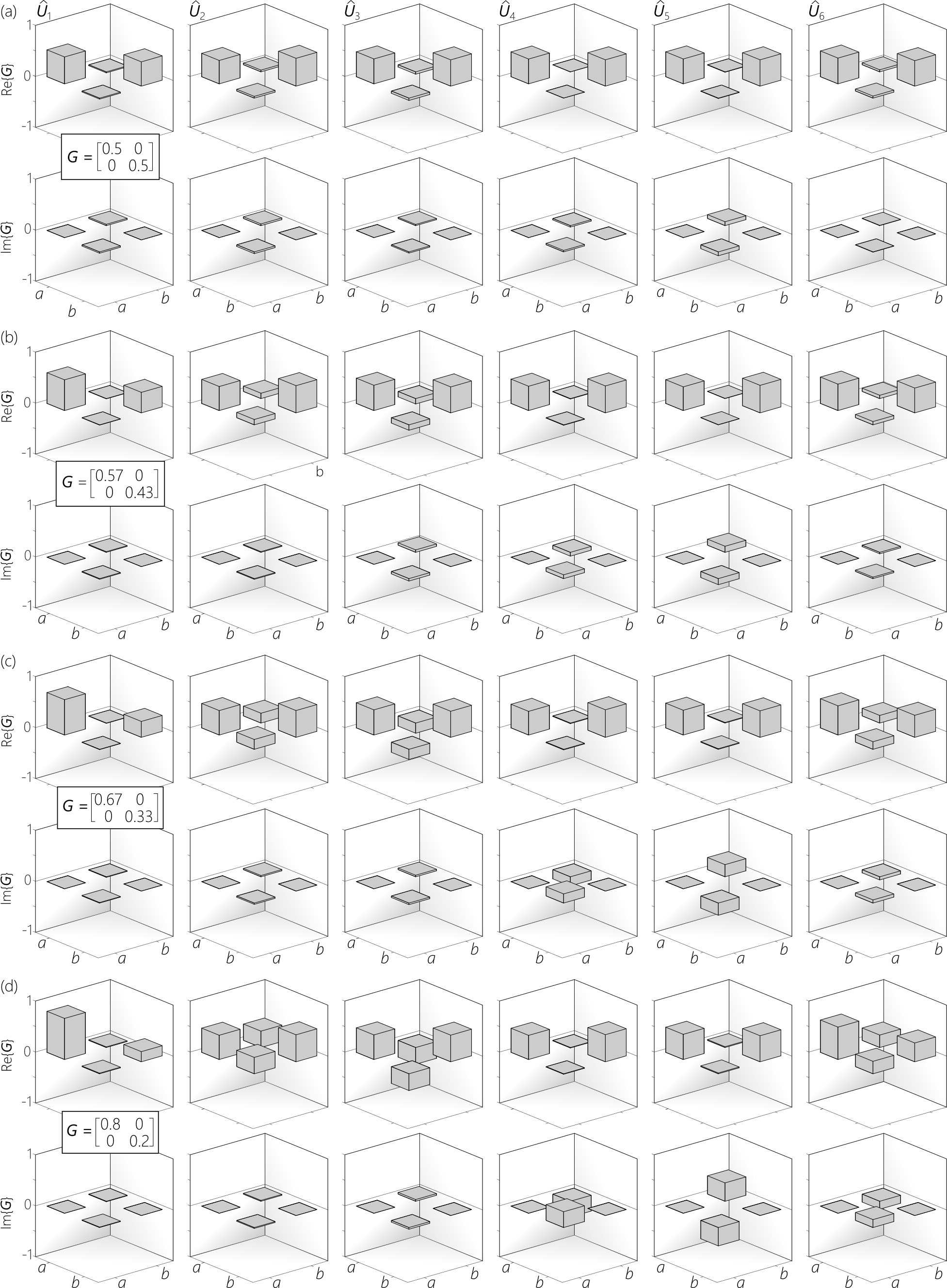} 
\caption{Reconstruction of the spatial coherence matrix $\mathbf{G}=\hat{U}\mathbf{G}^{\mathrm{D}}\hat{U}^{\dagger}$ for a partially coherent initial field after $\mathbf{G}^{\mathrm{D}}$ (left column) traverses $\hat{U}$. Each column shows the reconstructed $\mathbf{G}$ for the unitaries ($\hat{U}_{1}$ through $\hat{U}_{6}$) listed at the top (same as those in Fig.~\ref{fig:CoherentData}). We plot the real and imaginary parts of the coherence matrices, $\mathrm{Re}\{\mathbf{G}\}$ and $\mathrm{Im}\{\mathbf{G}\}$, respectively. (a) Incoherent field with $\lambda_{a}=\lambda_{b}=\tfrac{1}{2}$ and $D_{\mathrm{s}}=0$; (b) partially coherent field with $D_{\mathrm{s}}\approx0.143$; (c) $D_{\mathrm{s}}\approx0.333$; and (d) $D_{\mathrm{s}}=0.6$.}
\label{fig:PartiallyCoherentData}
\end{figure*}

To confirm the validity of our approach, we first carry out our measurements of the synthesis and reconstruction of prescribed coherence matrices using \textit{coherent} light. Measurements were carried out with both the on-chip laser source and the external laser (without the phase shifter and after blocking $|b\rangle$) with similar results. In both cases, the input field has the diagonal coherence matrix $\mathbf{G}^{\mathrm{D}}=\left(\begin{array}{cc}1&0\\0&0\end{array}\right)=|E\rangle\langle E|$, where $|E\rangle=\left(\begin{array}{cc}1\\0\end{array}\right)$ and $D_{\mathrm{s}}=1$, corresponding to optical power being delivered to a single input port. This field traverses the unitary $\hat{U}$, which takes on one of the following~6 forms:
\begin{eqnarray}\label{eq:6Unitaries}
\hat{U}_{1}\!\!\!\!&=&\!\!\!\!-e^{i\varphi}\left(\begin{array}{cc}1&0\\0&1\end{array}\right),\;\hat{U}_{2}=-\frac{1}{\sqrt{2}}e^{i\varphi}\left(\begin{array}{cc}1&-1\\1&1\end{array}\right),\nonumber\\
\hat{U}_{3}\!\!\!\!&=&\!\!\!\!-\frac{1}{\sqrt{2}}e^{i\varphi}\left(\begin{array}{cc}1&1\\-1&1\end{array}\right),\;\hat{U}_{4}=\frac{1}{\sqrt{2}}e^{i(\varphi+\pi/4)}\left(\begin{array}{cc}1&1\\i&-i\end{array}\right),\nonumber\\
\hat{U}_{5}\!\!\!\!&=&\!\!\!\!-\frac{1}{\sqrt{2}}e^{i(\varphi+\pi/4)}\left(\begin{array}{cc}1&1\\-i&i\end{array}\right),\nonumber\\
\hat{U}_{6}\!\!\!\!&=&\!\!\!\!\frac{1}{2}e^{i(\varphi+5\pi/12)}\left(\begin{array}{cc}\sqrt{3}&1\\e^{i\pi/6}&-\sqrt{3}e^{i\pi/6}\end{array}\right).
\end{eqnarray}
The unitary $\hat{U}_{1}$ is obtained from Eq.~\ref{eq:GeneralOnChip} after setting $\xi_{1}=\tfrac{\pi}{2}$, $\xi_{2}=-\tfrac{\pi}{2}$, and $\delta=\pi$; $\hat{U}_{2}$ with $\xi_{1}=\tfrac{\pi}{2}$, $\xi_{2}=\tfrac{\pi}{2}$, and $\delta=\tfrac{\pi}{2}$; $\hat{U}_{3}$ with $\xi_{1}=\tfrac{\pi}{2}$, $\xi_{2}=-\tfrac{\pi}{2}$, and $\delta=\tfrac{\pi}{2}$; $\hat{U}_{4}$ with $\xi_{1}=-\tfrac{\pi}{4}$, $\xi_{2}=\tfrac{\pi}{4}$, and $\delta=\tfrac{\pi}{2}$; $\hat{U}_{5}$ with $\xi_{1}=\tfrac{\pi}{4}$, $\xi_{2}=-\tfrac{\pi}{4}$, and $\delta=\tfrac{\pi}{2}$; and $\hat{U}_{6}$ with $\xi_{1}=-\tfrac{\pi}{12}$, $\xi_{2}=\tfrac{\pi}{12}$, and $\delta=\tfrac{2\pi}{3}$. Each of these unitaries transforms the generic initial coherence matrix $\mathbf{G}^{\mathrm{D}}$ into a new coherence matrix $\mathbf{G}=\hat{U}\mathbf{G}^{\mathrm{D}}\hat{U}^{\dagger}$.

We reconstruct $\mathbf{G}$ via measurements of the spatial Stokes parameters as outlined above. We plot in Fig.~\ref{fig:CoherentData} the~6 reconstructed coherence matrices, separating their real and imaginary parts, $\mathrm{Re}\{\mathbf{G}\}$ and $\mathrm{Im}\{\mathbf{G}\}$, respectively. To benchmark the performance of the unitaries, we borrow from quantum information processing the fidelity $F=(\mathrm{Tr}\{\sqrt{\mathbf{G}_{\mathrm{meas}}}\mathbf{G}_{\mathrm{theory}}\sqrt{\mathbf{G}_{\mathrm{meas}}}\})^{2}$ as a measure for the performance of $\hat{U}$ \cite{Jozsa94JMO}, where $\mathbf{G}_{\mathrm{theory}}$ and $\mathbf{G}_{\mathrm{meas}}$ are the theoretically expected and the measured coherence matrices, respectively. We find here that $F>0.98$, indicating excellent agreement between the measurements and theoretical expectations ($F=1$ indicates a perfect match), and also calculate $D_{\mathrm{s}}>0.97$ from $\mathbf{G}_{\mathrm{meas}}$. For $\hat{U}_{1}$, $\hat{U}_{2}$, and $\hat{U}_{3}$, $\mathrm{Im}\{\mathbf{G}\}=0$; for $\hat{U}_{4}$ and $\hat{U}_{5}$, the magnitude of the real and imaginary parts of the elements of $\mathbf{G}$ are equal; and for $\hat{U}_{6}$, the magnitudes of the real and imaginary parts are different. This confirms the ability to mold the field vector $|E\rangle$ for coherent light and the feasibility of on-chip coherence-matrix reconstruction.

\subsection{Measurements for a partially coherent field}

We next proceed to the modulation of partially coherent light. The external source delivers incoherent light to the chip described by the coherence matrix $\mathbf{G}=\tfrac{1}{2}\hat{\mathbb{I}}_{2}$ to paths $|a\rangle$ and $|b\rangle$. We then make use of an on-chip MZI to adjust the relative modal weights in the two paths to yield $\mathbf{G}^{\mathrm{D}}=\left(\begin{array}{cc}\lambda_{a}&0\\0&\lambda_{b}\end{array}\right)$. We adjust the ratio $\lambda_{a}:\lambda_{b}$ to $1:1$ corresponding to incoherent light with $D_{\mathrm{s}}=0$ [Fig.~\ref{fig:PartiallyCoherentData}(a)]; $1:0.75$ corresponding to $D_{\mathrm{s}}\approx0.143$ [Fig.~\ref{fig:PartiallyCoherentData}(b)]; $1:0.5$ corresponding to $D_{\mathrm{s}}\approx0.333$ [Fig.~\ref{fig:PartiallyCoherentData}(c)]; and $1:0.25$ corresponding to $D_{\mathrm{s}}=0.6$ [Fig.~\ref{fig:PartiallyCoherentData}(d)]. In each case, we reconstruct $\mathbf{G}$ via measurements of the spatial Stokes parameters and plot the real and imaginary parts of $\mathbf{G}$ separately. For each setting of $D_{\mathrm{s}}$, we implement the 6~unitaries given in Eq.~\ref{eq:6Unitaries}. For incoherent light [Fig.~\ref{fig:PartiallyCoherentData}(a)], where $\mathbf{G}=\tfrac{1}{2}\hat{\mathbb{I}}_{2}$, the unitaries leaves the coherence matrix invariant as expected. As $D_{\mathrm{s}}$ increases, the impact of any unitary is manifested more clearly. The fidelity for all cases is $F>0.98$, once again indicating excellent agreement between measurements and theoretical expectations.

\subsection{On-chip tuning of the interference visibility}

The measurements reported in Fig.~\ref{fig:CoherentData} and Fig.~\ref{fig:PartiallyCoherentData} make use of integrated on-chip detectors. However, it is critical to further confirm that the programmable on-chip system can deliver external optical fields with controllable coherence properties in concordance with the on-chip reconstructed coherence matrices. To this end, we rely on the general coherence matrix given in Eq.~\ref{eq:GeneralHermitian}, according to which the off-diagonal element is $G_{ab}=\tfrac{1}{2}e^{-i\alpha}D_{\mathrm{s}}\sin\delta$. The visibility of interference fringes resulting from superposing the modes $|a\rangle$ and $|b\rangle$ is $V=2|G_{ab}|=D_{\mathrm{s}}|\sin\delta|$. We tune $D_{\mathrm{s}}$ over the range $0\leq D_{\mathrm{s}}\leq1$ via the non-unitary operation preceding $\hat{U}$, and tune $\delta$ over the range $0\leq\delta\leq\tfrac{\pi}{2}$ ($0\leq\sin\delta\leq1$) via the unitary $\hat{U}$ [Fig.~\ref{fig:Chip}(d)]. Of course, when $D_{\mathrm{s}}=0$, the field is incoherent, and $\hat{U}$ has no impact, so that $V=0$ for all $\delta$. On the other extreme, when $D_{\mathrm{s}}=1$, the field is coherent, and $V=|\sin\delta|$, thereby reaching $V=1$ when $\delta=\tfrac{\pi}{2}$. This is to be expected because the degree of coherence is the maximum visibility observed once the field amplitudes are equalized (which occurs when $\delta=\tfrac{\pi}{2}$).

\begin{figure}[t!]
\centering
\includegraphics[width=8.7cm]{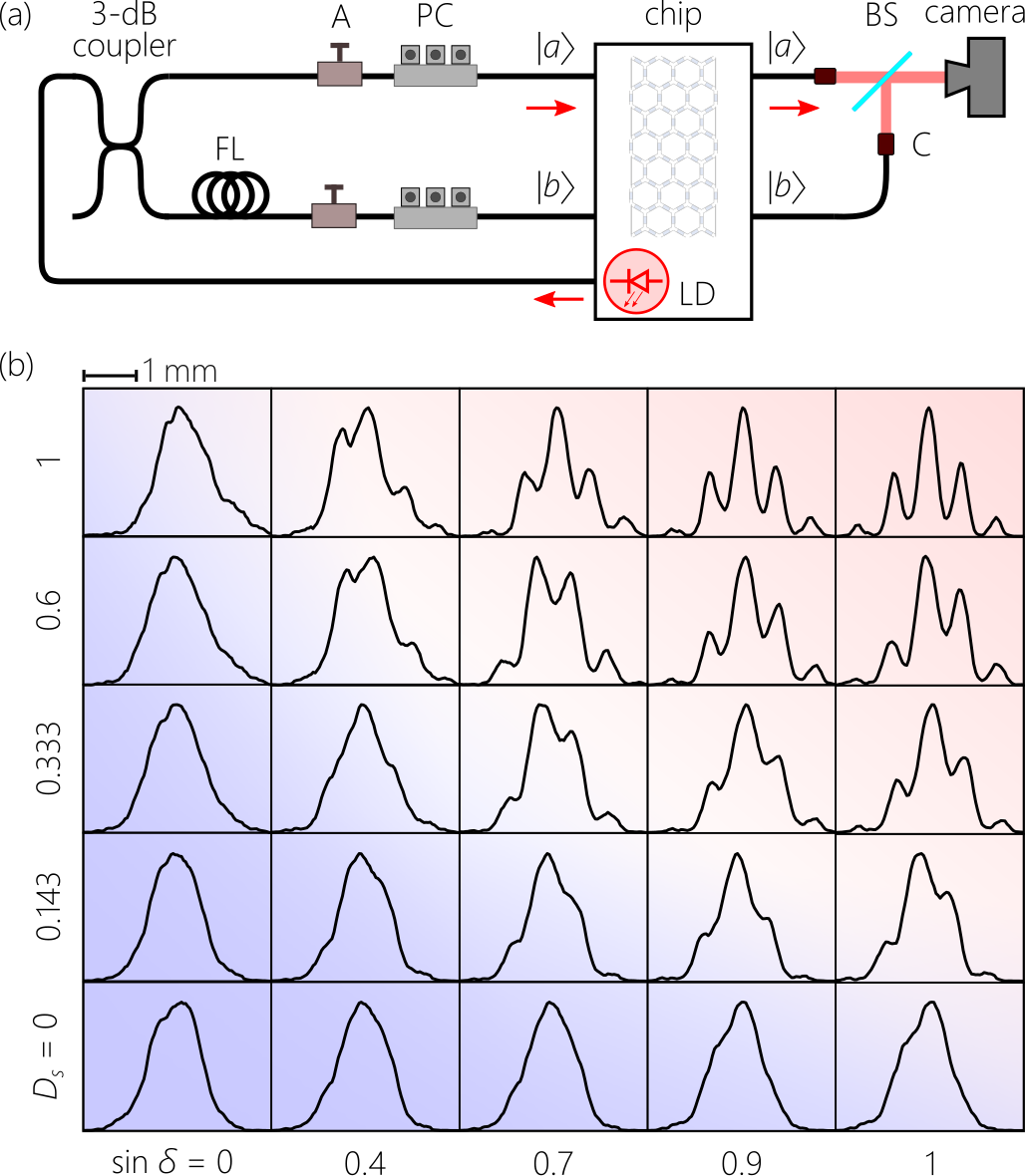} 
\caption{(a) Setup for on-chip tuning of the interference visibility observed in free space. LD: On-chip laser diode; FL: fiber loop; A: attenuator; PC: polarization controller; BS: symmetric beam splitter; C: fiber collimator. (b) One-dimensional sections through the interference patterns resulting from overlapping the fields associated with modes $|a\rangle$ and $|b\rangle$. The background shading indicates the direction of increasing $V$ (which increases with both $D_{\mathrm{s}}$ and $\sin\delta$).}
\label{fig:Visibility}
\end{figure}

In our measurements, we use the configuration depicted in Fig.~\ref{fig:Visibility}(a). We out-couple the on-chip laser diode (coherence length of a few meters) to an SMF split into two paths by a 3-dB coupler. We place a fiber spool (length of $\approx500$~m) in one path to delay the field beyond its coherence length. The two SMFs then couple the incoherent light to the chip as modes $|a\rangle$ and $|b\rangle$, as confirmed by the absence of any interference fringes when overlapping the two fields in free space, as done in Fig.~\ref{fig:Source}(c). This source allows us to reduce the camera integration time needed to record the interferogram is recorded after the chip [Fig.~\ref{fig:Visibility}(a)]. The fields associated with modes $|a\rangle$ and $|b\rangle$ for the synthesized $\mathbf{G}$ are out-coupled from the chip via SMFs ending in collimators, the two fields are then combined in free space at a symmetric beam splitter with a small relative angular tilt, and the intensity profile of their superposition is recorded by the camera. 

We plot in Fig.~\ref{fig:Visibility}(b) sections through the recorded interferograms. We note that the visibility increases as $\sin\delta$ increases and also as $D_{\mathrm{s}}$ increases (indeed, $V$ is their product). When $D_{\mathrm{s}}=0$, no fringes are observed at any setting. Moreover, for fixed $D_{\mathrm{s}}$, the maximum visibility is observed when $\sin\delta\rightarrow1$.

\begin{figure*}[t!]
\centering
\includegraphics[width=16.3cm]{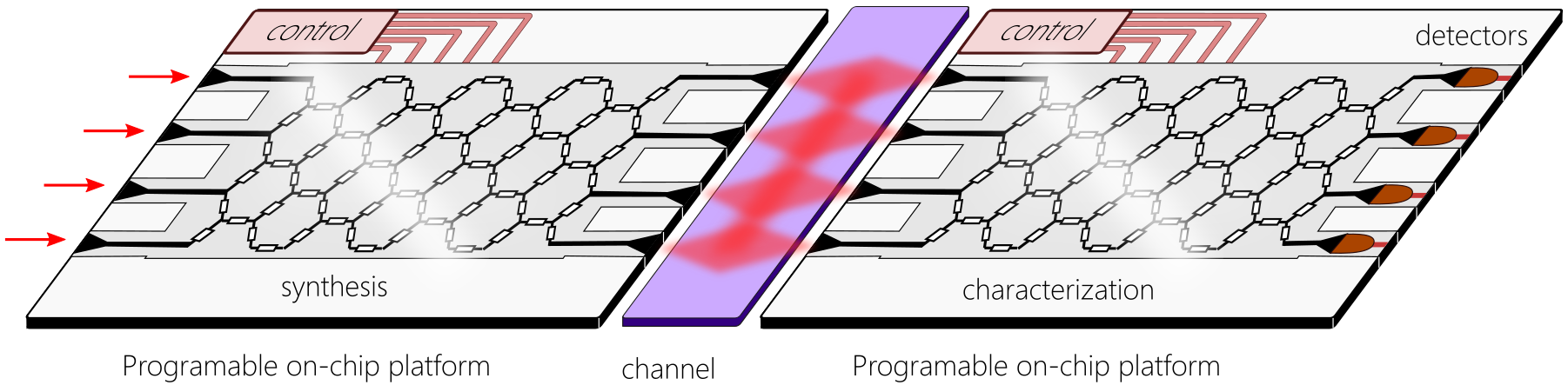} 
\caption{Vision for programmable on-chip manipulation of partial optical coherence. We consider an optical field characterized by a finite number of modes $N$ in a discrete basis, which we term `structured coherence'. The envisioned system  comprises two programmable on-chip platforms. Generic incoherent light is coupled to the first chip on the left. This `synthesis' chip tunes the degree of coherence and implements an arbitrary unitary $\hat{U}$ to realize a prescribed coherence matrix $\mathbf{G}$. After traversing a `channel' (transmission across a multimode fiber, scattering from an object, etc.), a second `characterization' chip receives the coherence matrix $\mathbf{G}'$ and reconstructs it with on-chip detectors.}
\label{fig:Vision}
\end{figure*}

\section{Discussion and Conclusions}

It has long been established that an $N\times N$ unitary operating on $N$ modes \cite{Reck94PRL,Bogaerts20Nature,Saleh25book} can be decomposed into a sequence of $2\times2$ unitaries operating on pairs of modes, limited by the number of available MZI's in the on-chip platform. Consequently, the unitaries reported here are the basic building blocks for future arbitrary unitaries. Moreover, the reconstruction of coherence matrices can be achieved through extensions to the Stokes-parameters strategy employed here \cite{Kagalwala15SR}.

Our results therefore lay the foundation for realizing the vision sketched in Fig.~\ref{fig:Vision} to fully exploit the `coherence advantage'. Generic incoherent light is delivered to an optical chip (or the incoherent light may be produced on the chip), where the degree of coherence is tuned and the initial diagonal coherence matrix $\mathbf{G}^{\mathrm{D}}$ is unitarily transformed into a prescribed coherence matrix $\mathbf{G}$. The produced field characterized by this coherence matrix is then launched into an optical channel (e.g., free-space propagation, a multimode fiber, a bundle of single-mode fibers, scattering from an object or from a medium, or possibly a different chip). The field emerging from the optical channel is then coupled to a second programmable chip after separating out the modes into different waveguides. The coherence matrix of the received field is reconstructed on the second chip, enabled by integrated on-chip detectors. This procedure can be adaptable, responding dynamically to external stimuli, coupled potentially with on-chip feedback and/or feedforward. 

The results reported here may find applications in several areas. For example, a tunable partially coherent source can be utilized in variable coherence tomography \cite{Baleine04OL,Baleine04JOSAA,Baleine05PRL}. Once the unitaries are extended to $N\times N$ matrices, such a chip can be utilized as a programmable transmitter and receiver for coherence-rank communications across a scattering channel \cite{Harling25APLP}, or for utilizing modal correlations for high-channel-capacity communications over a multimode fiber \cite{Nardi22OL}. 

Several challenges need to be addressed along the way to constructing a real-world implementation of on-chip programming of partially coherent light. First, efficient sources of SMF-coupled partially coherent light with large modal dimensions are needed. In free space, rotating ground glass is commonly used to reduce the spatial coherence of an initially coherent wave front. The efficiency of such an approach is extremely low if the field is to be coupled into SMFs, so new approaches need to be pursued. Second, monitoring the optical polarization is critical in strongly scattering channels, yet almost all on-chip platforms make use of a single polarization component. It is thus imperative to couple polarization information to the chip, possibly by converting the polarization information first to spatial information \cite{Abouraddy17OE,Okoro17Optica}. Third, care needs to be taken for large on-chip systems to maintain the maximum length differences between the paths followed by the modes to be smaller than the coherence length of the source. Fourth, as the dimension of $\mathbf{G}$ increases, the constraints on the steps (and thus the requisite time) needed for its reconstruction become pressing, especially when used in a high-speed communications link.

In conclusion, we have reported the first results on the synthesis and characterization of two-mode partially coherent optical fields in a programmable on-chip platform. Starting with a generic incoherent two-mode field, we utilize the on-chip system to construct non-unitary transformations that tune the degree of coherence, then unitarily synthesize arbitrary $2\times2$ coherence matrices, before reconstructing them via measurements of the spatial Stokes parameters. These results open up a new path for fundamental investigations of partial optical coherence of large-dimensional fields and for exploiting the coherence advantage in new applications in optical communications, cryptography, computation, and spectroscopy.


Funding: 
U.S. Office of Naval Research (ONR) N00014-20-1-2789.

Acknowledgment:
We thank Guifang Li and Midya Parto for lending equipment.


\bibliography{diffraction}

\end{document}